\documentclass[namedreferences]{SolarPhysics}
\usepackage[optionalrh]{spr-sola-addons} 
\usepackage{graphicx}
\usepackage{color}           
\usepackage{url}             

\newcommand{\adv}{    {\it Adv. Space Res.}}

\newcommand{\aap}{    {\it Astron. Astrophys.}}

\newcommand{\ag}{     {\it Ann. Geophys.}}

\newcommand{\apj}{    {\it Astrophys. J.}}
\newcommand{\apjl}{   {\it Astrophys. J. Lett.}}

\newcommand{\grl}{    {\it Geophys. Res. Lett.}}

\newcommand{\jgr}{    {\it J. Geophys. Res.}}

\newcommand{\solphys}{{\it Solar Phys.}}

\newcommand{\ssr}{    {\it Space Sci. Rev.}}

\begin{document}

\begin{article}

\begin{opening}

\title{Magnetic Flux of EUV Arcade and Dimming Regions as a Relevant Parameter
for Early Diagnostics of Solar Eruptions -- Sources of
Non-Recurrent Geomagnetic Storms and Forbush Decreases}

\author{I.M.~\surname{Chertok}$^{1}$\sep
    V.V.~\surname{Grechnev}$^{2}$\sep
        A.V.~\surname{Belov}$^{1}$\sep
        A.A.~\surname{Abunin}$^{1}$
        }

\runningauthor{Chertok et al.}
 \runningtitle{EUV/magnetic Diagnostics of Solar Eruptions}

\institute{${}^{1}$Pushkov Institute of Terrestrial Magnetism,
            Ionosphere and Radio Wave Propagation (IZMIRAN), Troitsk, Moscow
            Region, 142190 Russia email: \url{ichertok@izmiran.ru}\\
            ${}^{2}$Institute of Solar-Terrestrial Physics SB RAS,
            Lermontov St.\ 126A, Irkutsk 664033, Russia email: \url{grechnev@iszf.irk.ru}
}

\begin{abstract}
This study aims at the early diagnostics of geoeffectiveness of
coronal mass ejections (CMEs) from quantitative parameters of the
accompanying EUV dimming and arcade events. We study events of the
23th solar cycle, in which major non-recurrent geomagnetic storms
(GMS) with Dst $<-100$ nT are sufficiently reliably identified
with their solar sources in the central part of the disk. Using
the SOHO/EIT 195~\AA\ images and MDI magnetograms, we select
significant dimming and arcade areas and calculate summarized
unsigned magnetic fluxes in these regions at the photospheric
level. The high relevance of this eruption parameter is displayed
by its pronounced correlation with the Forbush decrease (FD)
magnitude, which, unlike GMSs, does not depend on the sign of the
$B_z$ component but is determined by global characteristics of
ICMEs. Correlations with the same magnetic flux in the solar
source region are found for the GMS intensity (at the first step,
without taking into account factors determining the $B_z$
component near the Earth), as well as for the temporal intervals
between the solar eruptions and the GMS onset and peak times. The
larger the magnetic flux, the stronger the FD and GMS intensities
are and the shorter the ICME transit time is. The revealed
correlations indicate that the main quantitative characteristics
of major non-recurrent space weather disturbances are largely
determined by measurable parameters of solar eruptions, in
particular, by the magnetic flux in dimming areas and arcades, and
can be tentatively estimated in advance with a lead time from 1 to
4 days. For GMS intensity, the revealed dependencies allow one to
estimate a possible value, which can be expected if the $B_z$
component is negative.
\end{abstract}


\end{opening}


\section{Introduction}
 \label{S-introduction}

Coronal mass ejections (CMEs) are the most grandiose manifestation
of the solar activity in terms of their size, energy, and space
weather effects (\textit{e.g.}, \opencite{Kunow2006};
\opencite{Gopalswamy2010}; and references therein). They are
connected with large-scale magnetic rearrangements in the solar
atmosphere and expel a bulk of magnetized plasma into the
interplanetary space. CMEs and their interplanetary counterparts
ICMEs are prime drivers of the most severe non-recurrent space
weather disturbances, in particular such important and strongly
effective ones as major geomagnetic storms (GMSs)
\cite{Gosling1993,BothmerZhukov2007,Gopalswamy2009}. The latter
occur when large and fast CMEs erupt mainly from the central
region of the visible solar disk as a partial or full halo CME and
the corresponding ICMEs bring to the Earth a sufficiently strong
and prolonged southward (negative) magnetic field $B_z$ component
either in the flux rope or in the sheath between the flux rope and
the ICME-driven shock. Simultaneously the magnetized ICMEs deflect
galactic cosmic rays entering the heliosphere and cause reduction
of their intensity measured at the Earth and in the near-Earth
space called non-recurrent Forbush decreases (FDs)
\cite{Cane2000,Belov2009,RichardsonCane2011}. There are also
generally less intense recurrent GMSs and FDs caused by corotating
interaction regions (CIRs) which are formed as a result of
interaction between the fast solar wind from coronal holes and the
preceding slow wind from closed magnetic structures
\cite{Richardson2006,Zhang2007a}. We will concentrate below just
on the non-recurrent GMSs and FDs leaving the recurrent ones
beyond the scope of our consideration.

\subsection{Existing Diagnostic Methods}
One of the most important tasks of the solar-terrestrial physics
and space weather prediction is diagnostics of geoeffectiveness of
CMEs, \textit{i.e.}, quantitative forecast of a possible
non-recurrent GMSs and FDs from observed characteristics of the
eruption that just occurred. Existing algorithms of such
diagnostics are based in one way or another on the measurements of
the CME speed and shape in the plane of the sky in the near-Sun
region from the data of SOHO/LASCO \cite{Brueckner1995}. A number
of direct empirical relations have been established between the
projected or deprojected CME expansion speed and transit time,
\textit{i.e.}, an interval between the moments of a CME eruption
from the Sun and ICME arrival to 1 AU
\cite{Gopalswamy2001,SiscoeSchwenn2006,Xie2006,Kim2007,GopalswamyXie2008,Michalek2008}.
As for the GMS intensity, it strongly depends on the magnetic
field strength and orientation in the corresponding ICME. The
required presence of the southern $B_z$ component can be generally
determined from the orientation of the magnetic field in the CME
source region, from the shape (S or inverse S) of the pre-eruption
X-ray sigmoid, from the orientation angle of elongated LASCO CME
and post-eruption arcade, as well as from the local tilt of the
coronal neutral line at 2.5 solar radii
\cite{Kang2006,Song2006,YurchyshynTripathi2009}.

In empirical algorithms for the forecast of the GMS intensity, the
same near-Sun CME speeds are also used as one of the main input
parameters. The corresponding algorithms are also combinations of
several methods. For example, in the algorithm of
\inlinecite{Yurchyshyn2004} (see also \opencite{Yurchyshyn2005}),
the expected magnitude of the $B_z$ component in ICMEs near the
Earth is firstly estimated by established correlation with the
projected CME speed, and then a statistically revealed
relationship between $B_z$ and the Dst geomagnetic index is used.
Recently \inlinecite{Kim2010} presented empirical expressions for
the Dst index calculated from the plane-of-the-sky CME speed,
direction parameter, and (heliographic) longitude for two CME
groups depending on whether the magnetic fields are oriented
southward or northward in their source regions.

In addition to the empirical/statistical tools, some analytical
models and numerical MHD simulations have been developed
particularly for forecasting of the ICME arrival time at 1~AU
(\textit{e.g.}, \opencite{SiscoeSchwenn2006};
\opencite{Smith2009}; \opencite{Taktakishvili2009};
\opencite{Vrsnak2010}; and references therein). Again, the
near-Sun CME characteristics and some additional data are used as
input parameters in the models describing the ICME-driven shock
propagation in the solar wind taking into account the
`aerodynamic' drag, interaction with CIRs, and other effects.

With the advent of the STEREO era \cite{Kaiser2008} it has become
possible to trace propagation of the Earth-directed CMEs in the
corona and ICMEs in the interplanetary space from three vantage
points simultaneously (with two STEREO and SOHO spacecraft) and to
use stereoscopic methods for reconstruction of the 3D trajectory,
angular width, and speed of the corresponding ICMEs
(\textit{e.g.}, \opencite{Liu2010}; \opencite{Lugaz2010};
\opencite{Wood2011}). Valuable information about ICMEs and their
geoefficiency is also obtained from multi-point interplanetary
scintillation (IPS) radio measurements and observations with the
\textit{Solar Mass Ejection Imager} \cite{Jackson2004}, especially
in combination with the SOHO and twin-spacecraft STEREO data
(\textit{e.g.}, \opencite{Jackson2009}; \opencite{Webb2009};
\opencite{Manoharan2010}). Nevertheless, diagnostics of CMEs from
observations of their low-corona signatures remains a very urgent
topic, because it can provide the earliest alert on
geoeffectiveness of solar eruptions.

\subsection{Background for EUV/Magnetic Diagnostics}

In this paper, we present a new approach to the early diagnostics
of solar eruptions in which quantitative characteristics of such
large-scale CME manifestations as dimming and formation of
post-eruption (PE) arcades observed in the extreme ultraviolet
(EUV) range are used as key parameters instead of the projected
CME speed and shape. The idea of such an approach was proposed by
\inlinecite{ChertokGrechnev2006}. The total (unsigned) magnetic
flux of the longitudinal field at the photospheric level within
the dimming and arcade areas is considered as a main quantitative
parameter of eruptions. The magnetic flux of a CME can be possibly
somewhat less than the whole magnetic flux in dimming and arcade
regions (see, \textit{e.g.}, \opencite{GibsonFan2008}). However,
such a total unsigned flux can actually serve as a measure of the
erupting flux. For simplicity, we will call the total unsigned
magnetic flux in dimming and arcade areas at the photospheric
level the `eruptive magnetic flux' or `eruption parameter'.

Dimmings are CME-associated regions in which the EUV (and soft
X-ray as well) brightness of coronal structures is temporarily
reduced during an ejection and persists over many hours. Deep and
extended core dimmings are formed near the center of an eruption,
and additional remote dimmings can also be observed at a large
part of the solar surface
\cite{Thompson1998,HudsonCliver2001,Harra2011}. The deepest
stationary long-lived dimmings adjacent to the eruption center are
interpreted mainly as a result of plasma outflow from the
footpoints of erupting and expanding CME flux ropes
\cite{Sterling1997,Webb2000}. It is noteworthy that, as
near-the-limb eruptions reveal, the extent of the dimming area
corresponds to the apparent angular size of the corresponding CME
observed with white-light coronagraphs \cite{Thompson2000}.

Large-scale arcades of bright loops enlarging in size over time
arise at the place of the main body of pre-eruption magnetic flux
ropes ejected as CMEs
\cite{Kahler1977,Sterling2000,HudsonCliver2001,Tripathi2004}. Such
arcades with extended emitting ribbons in their bases are formed
in active regions above magnetic neutral lines under erupting
magnetic flux ropes, which then develop into CMEs. While the core
dimmings correspond to footpoints of the erupted flux ropes, the
PE arcades can be considered as counterparts of the central
flaring part of these flux ropes. As a whole, dimmings and PE
arcades visualize structures and areas involved in the process of
the CME eruption. This gives reasons to expect that their
quantitative parameters, in particular magnetic fluxes, can be
relevant and promising for early quantitative evaluations of
geoeffectiveness of the corresponding ICMEs.

\begin{figure} 
  \centerline{\includegraphics[width=\textwidth]
   {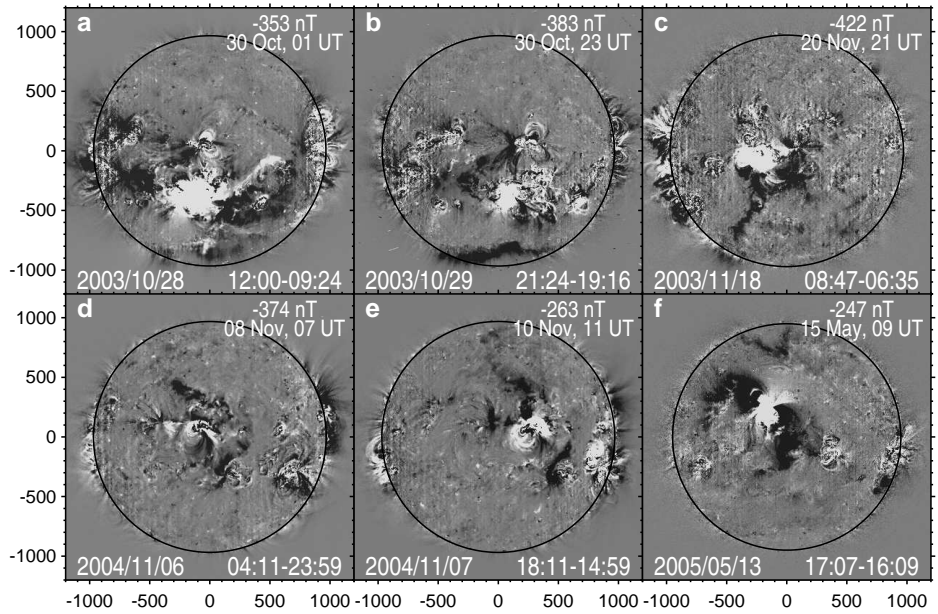}
  }
\caption{Dark dimmings and bright post-eruption arcades in solar
sources of the severest GMSs over the solar cycle 23 in SOHO/EIT
195~\AA\ fixed-base difference images. The Dst of the strongest
disturbances and the dates and time of their registration are
specified in the upper right corner of each panel. The dates and
time of EIT images subjected to the subtraction are specified at
the bottom of each panel.
    }
  \label{F-fig1}
  \end{figure}

Figure~\ref{F-fig1} shows the dimmings and PE arcades accompanying
the eruptions which were the sources of the strongest GMSs of the
solar cycle 23, as they look like in the derotated difference
images of the EUV telescope SOHO/EIT \cite{Delab1995} in the
195~\AA\ channel. This figure illustrates that large eruptions can
be global in nature and probably involve octopus-like bundles of
magnetic ropes anchored in several interconnected active regions
\cite{ChertokGrechnev2005,Zhang2007c,ZhukovVeselovsky2007}.

Until now in the space weather aspect, the qualitative information
about dimmings and arcades was mainly used as a tool for
identification of frontside CMEs originating in eruptions on the
visible solar disk and to distinguish them from backside halo CMEs
\cite{Zhukov2005}. Further, the majority of full and partial halo
CMEs are elongated in the direction of the axial field of the PE
arcades that can give an indication in advance about the sign of
the $B_z$ component in the associated ICMEs near the Earth
\cite{YurchyshynTripathi2009}. As for quantitative parameters,
there are some studies in which the photospheric magnetic flux in
dimming regions was calculated for limited samples of events and
compared with the model magnetic flux in ICMEs, particularly in
their magnetic cloud variety, computed from \textit{in situ}
measurements at 1~AU (see \opencite{Demoulin2008};
\opencite{Mandrini2009} for a review).

\subsection{Outline}
The main points of our approach and their presentation in this
paper are as follows:
\begin{itemize}
 \item
 The eruptions from the central zone of the disk occurring
throughout the solar cycle 23 responsible for major non-recurrent
GMSs of the disturbance storm time index Dst $<-100$~nT, are
considered.
 \item
The photospheric magnetic fluxes not only in dimming, but also in
PE arcades are considered. This is in agreement with the
conclusion of \inlinecite{Mandrini2007} and \inlinecite{Qiu2007}
that the magnetic flux in the dimming area only is not sufficient
to account for the observed ICME flux. The data and selection
method of the dimming and arcade areas as well as the computation
procedure of the summarized unsigned magnetic flux within them as
a measure of the ejected CME flux are described in
Section~\ref{S-data_technique}.
 \item
 Bearing in mind the forecast as a result of the analysis, the eruptive
fluxes are correlated directly with the magnitude of FDs,
intensity of GMSs and transit times, omitting comparison with the
ICME parameters near the Earth.
 \item
To test the efficiency of our approach, it is reasonable to begin
with a correlation between eruptive magnetic fluxes and FD
magnitudes (Section~\ref{S-forbush_decreases}) since the intensity
of GMSs strongly depends on the $B_z$ component in a relatively
local ICME part interacting directly with the Earth's
magnetosphere, while the FD magnitude does not depend on $B_z$ and
is determined by the magnetic field strength in a global ICME, as
well as by its speed and sizes (\textit{e.g.},
\opencite{Belov2009}).
 \item
 Positive results obtained for FDs show relevance of the
eruptive flux as a diagnostic parameter and encourage its similar
comparison with intensity of GMSs, while without taking into
account factors determining the $B_z$ component
(Section~\ref{S-geomagnetic_storms}).
 \item
 The analysis shows that not only the magnitudes of FD and GMS are
closely related to the eruptive magnetic flux, but that the latter
largely determines also the times of ICME propagation from the Sun
to the Earth (Section~\ref{S-transit_times}). This is true for two
transit times that we consider, both of which are measured from
the eruption moment at the Sun: (a)~the onset transit time,
\textit{i.e.}, an interval until the interplanetary disturbance
arrival at the Earth, and (b)~the peak transit time,
\textit{i.e.}, an interval until the GMS peak.
 \item
 Summary and discussion, including some results of the testing
of the obtained tentative relations by their application to actual
data of 2010 and directions of the relevant further
investigations, are given in Section~\ref{S-discussion}.
\end{itemize}
 Some preliminary results concerning FDs and the ICME onset transit
time were published in a brief paper of
\inlinecite{ChertokBelovGrechnev2011}.

\section{Data and Technique}
 \label{S-data_technique}

\subsection{Events}

Our analysis is based on the catalog of major GMSs prepared by the
Living with a Star (LWS) Coordinated Data Analysis Workshop (CDAW;
\citeauthor{Zhang2007a} \citeyear{Zhang2007a,Zhang2007b}). The
catalog contains data on the most intense GMSs with a minimum Dst
$< -100$ nT that occurred during 1996\,--\,2005 including data on
their solar and interplanetary sources. Additionally, we took into
account the revised and updated ICME list of cycle 23 compiled by
\inlinecite{RichardsonCane2010} and containing information on
their probable solar sources, basic properties, and geomagnetic
effects.

In the CDAW catalog \cite{Zhang2007a}, all events are classified
into three types depending on the character of GMS, its
interplanetary drivers, and solar sources: (a)~S-type, in which
the separate storm is associated with a single ICME and a single
eruption (CME) at the Sun; (b)~M-type, in which the compound storm
is associated with multiple, complex, probably interacting ICMEs
arising from multiple solar eruptions and CMEs; (c)~C-type, in
which GMSs are associated with the solar wind CIRs caused by
high-speed streams from coronal holes. Three confidence levels of
the GMS identification with solar sources are distinguished: the
highest confidence level 1 means a clear unambiguous
identification with a concrete source at the Sun; the less
confidence level 2 denotes a less reliable but probable
identification with more than one source; the low confidence level
3 belongs to an ambiguous identification and problematic events.
M-type GMSs automatically fall into levels 2 or 3 because of their
intrinsic complexity. For M-type GMSs, in the cases where this was
possible, we extracted the strongest decrease of the Dst index,
and the most powerful solar eruption (producing a strongest flare
and most energetic CME) identified according to the corresponding
dimming events and PE arcades that occurred at a suitable time was
considered as its most probable source with an ambiguous
identification level. For such events only these eruptions are
presented in our Table~\ref{T-table} (see below). In
Table~\ref{T-table} and in the text to follow, the date is
expressed for simplicity as year/month/day.

We only deal with non-recurrent GMSs of types S and M initiated by
sporadic solar eruptions and CMEs. Therefore, C-type events
associated with coronal holes are omitted entirely. Moreover, to
minimize the projection effect on the dimming and arcade
parameters, we considered GMSs identified with eruptions which
occurred in the central zone of the visible solar hemisphere
within $\pm 45^\circ$ from the disk center. It would be more
reasonable to use the  $\pm 30^\circ$ limit \cite{Wang2002}, but
in this case the number of analyzed events were significantly
reduced. Several non-recurrent and intense GMS events, including
those from central solar sources, were removed from our
consideration due to data gaps either of the whole SOHO spacecraft
(CDAW storm No.\,6, 1997/11/23; No.\,11, 1998/08/06; No.\,13,
1998/08/27; No.\,14, 1998/09/25; No.\,20, 1999/02/18; and No.\,23,
1999/11/13) or the absence of EIT images (No.\,47, 2001/11/06;
No.\,72, 2004/04/04; and No.\,85, 2005/06/12), or the absence of
the SOHO/MDI \cite{Scherrer1995} magnetograms (No.\,15,
1998/10/19)\footnote{Here and afterwards the GMS events are
numbered according to the CDAW catalog (\citeauthor{Zhang2007a}
\citeyear{Zhang2007a,Zhang2007b})}. Of course, we did not analyze
GMSs whose solar source is unknown, for example such as CDAW storm
No.\,2, 1997/04/22; No.\,7, 1998/02/18;  No.\,28, 2000/08/11;
No.\,31, 2000/10/05; No.\,40, 2001/04/22; and No.\,58, 2002/10/01.

In the course of consideration, we carried out verification and
some corrections of the CDAW identification of GMSs with
corresponding solar eruptions paying particular attention to the
characteristics of appropriate dimmings and PE arcades and taking
into account the data base created in IZMIRAN \cite{Belov2009},
the online SOHO/LASCO CME catalog
(\url{http://cdaw.gsfc.nasa.gov/CME_list/};
\opencite{Yashiro2004}), as well as all accessible solar and
solar-terrestrial data acquired by ground-based and space-borne
observatories.In particular, several S-type events from the CDAW
catalog with a `source unknown' classification were reconsidered
and suitable sources of the identification level 2 were determined
for these events:
 \begin{itemize}
 \item
 For example, for the S-type storm No.\,34 (2000/11/06) the `source
unknown' proposition concerns a flare only. However, there was a
suitable major halo CME observed on 200/11/03 after
18:26\footnote{All times hereafter are UT.} that was accompanied
by a large PE arcade and noticeable dimmings near the solar disk
center.
 \item
 Another S-type GMS No.\,36 (2001/03/20) of this kind was certainly
caused by eruptions from the central active region (AR) 9373 and
its surroundings on 2001/03/15 and 2001/03/16. Judging by surface
activities visible in the EIT images, the eruption of 2001/03/15,
21:00 was accompanied by large dimmings and PE arcade and could
give the main contribution to this GMS.
 \item
 Similarly, for the S-type GMS No.\,76 (2004/08/30) instead of the
unknown source we accept a filament eruption of 2004/08/26, 12:00
near AR 10664 (S11W38) with apparent EIT signatures and a large,
slowly accelerating CME as a probable source.
 \end{itemize}
 Among our other refinements of the CDAW catalog, the following
ones should be also mentioned:
 \begin{itemize}
 \item
We merged two storms No.\,16 (1998/11/08) and No.\,17 (1998/11/09)
in the catalog into one event (No.\,17) of the identification
level 1, because this disturbance was caused by different parts of
a single ICME resulting mainly from the solar eruption on
1998/11/05, 19:55.
 \item
The catalog unambiguously associates the single GMS No.\,24
(2000/02/12) with a central eruption and a halo CME of 2000/02/10,
02:30. However, another much more powerful eruption with a
spectacular halo CME, large dimmings, and PE arcade occurred on
2000/02/09, 20:00 in AR 8853 (S17W40). For this reason, the
eruption of 2000/02/09 is considered as a basic source of the
storm with the identification level~1.
 \item
In the catalog, a strong halo CME of 2003/08/14, 20:06 is
indicated as a possible source of the single GMS 66 (2003/08/18)
but with the identification level 2 because dimmings were not
detected in the central part of the disk in connection with this
CME. Our processing of EIT data revealed that not only significant
dimmings but also a PE arcade near AR 10431 accompanied this CME.
Consequently, the identification level can be raised to~1.
 \item
According to the catalog, the S-type GMS No.\,74 (2004/07/25) is
identified unambiguously with an eruption of 2004/07/22 at
$\approx\,$08:30. Meanwhile, judging by parameters of the dimming
and PE arcade, a more powerful central trans-equatorial eruption
occurred on this day at 22:58. Therefore, it is reasonable to
consider the later eruption as a probable source of this storm
with the confidence level~2.
 \item
The catalog indicates that the great GMS No.\,77 (2004/11/08) was
probably caused by two solar eruptions of 2004/11/04 accompanied
by CMEs and the C6.3 and M5.4 flares, which peaked at 09:05 and
22:29. In our opinion, the main contribution to this storm,
including its sudden commencement on 2004/11/07, 18:27, was
provided by a more powerful eruption of 2004/11/06 with a halo CME
and M9.3 flare at 00:34. On the other hand, another powerful
eruption of 2004/11/07 associated with a halo CME and X2.0 flare
at 16:06 is considered by us as a main source of the subsequent
great GMS No.\,78 (2004/11/10). Bearing in mind the discrepancies
with the CDAW catalog, these two events are classified further as
multiple storms with confidence level 2 of the source
identification.
 \item
For the multiple event No.\,79 (2005/01/18), the catalog refers to
two eruptions of 2005/01/15 as probable sources. It seems more
probable that these and some earlier eruptions were responsible
for the initial GMS disturbances starting on 2005/01/15, but the
main Dst decrease on 2005/01/18 most likely was caused by a
powerful eruption of 2005/01/17, 09:52 characterized by one of the
fastest CME as well as by large dimmings and arcade.
 \item
The catalog classifies GMS No.\,81 (2005/05/08) as a coronal hole
(CIR) associated one. Meanwhile, a spectacular large-scale
eruption, which occurred on 2005/05/06 at $\approx\,$17:00 around
AR~0758 (S09E28) and was accompanied by a fast halo CME and a C8.5
long-duration flare, can be considered as a probable source of
this storm with the confidence level~2.
 \end{itemize}
It is important that reconsideration of all the events listed in
the two last paragraphs resulted in reasonable transit times of
the corresponding ICMEs to the Earth (see
Section~\ref{S-transit_times}). To encompass the whole cycle 23,
two strong GMSs with Dst $<-100$ nT on 2006/04/14 and 2006/12/15
should be mentioned. The first storm was excluded from our
analysis because its solar source is unknown, and the second one
was added into our Table~\ref{T-table} as event 90.

Among non-recurrent GMSs, we discriminate events initiated by
eruptions occurring in ARs and events associated with filament
eruptions outside ARs (the latter are marked in the CDAW catalog
with a note `QS', \textit{i.e.}, a quiet-Sun region). The reasons
are that these two categories of eruptions differ significantly in
characteristics of accompanying dimmings and PE arcades,
properties of CMEs/ICMEs, and intensity of GMSs and FDs, which
they cause (see, \textit{e.g.}, \opencite{Svestka2001};
\opencite{Chertok2009}; \opencite{Gopalswamy2009}). For brevity we
refer them to as AR events and non-AR events, respectively.

\subsection{Analyzed Parameters}

As a measure of the GMS intensity, we use the minimum final hourly
Dst index for all events of 1997\,--\,2006
(\url{http://wdc.kugi.kyoto-u.ac.jp/dstdir/index.html}). In the
CDAW catalog for events of 2004\,--\,2006 the provisional Dst
values were used. Now in events No.\,73 (2004/07/23) and No.\,83
(2005/05/20) the absolute value of the final Dst index is slightly
less than 100~nT. Nevertheless, these two single events are kept
in our set because they have the highest identification level S1.

As for a FD characteristic, its maximum magnitude is adopted which
corresponds to a cosmic ray rigidity of 10 GV and is determined
from data of the world network of neutron monitors using the
global survey method \cite{Krymskii1981,Belov2005}. In some
complex events, a secondary significant FD was observed against
the strong background of the descending phase of a previous strong
FD. This occurred, for example, in the CDAW paired events No.\,50
and No.\,51 (2002/04/18\,--\,20) and No.\,67 and No.\,68
(2003/10/30). In such cases, we considered and included in
Table~\ref{T-table} the magnitude of the first FD only, because
the true value of the secondary FD is difficult to determine due
to several factors influencing in this complex situation the
measured cosmic ray intensity.

In considering the temporal parameters of GMSs, the peak time of
the corresponding soft X-ray flare (see
\url{http://www.swpc.noaa.gov/ftpmenu/warehouse.html}) was taken
as an eruption time at the Sun. For a couple of events initiated
by filament eruptions outside ARs that were not accompanied by a
noticeable soft X-ray flare, the eruption time was taken to be
equal to the peak emission time of a PE arcade visible in EIT
195~\AA\ images, as described below.

In this study, we analyze two transit times, which adequately
characterize GMSs and are important for their forecasting. The
onset transit time ($\Delta T_0$) is defined as an interval
between the eruption time (the peak time of an associated soft
X-ray burst) and the arrival time of the corresponding
interplanetary disturbance (shock wave) to the Earth which is
indicated particularly by the geomagnetic storm sudden
commencement (SSC)
(\url{ftp://ftp.ngdc.noaa.gov/STP/SOLAR_DATA/SUDDEN_COMMENCEMENTS/STORM2.SSC}).
The peak transit time ($\Delta T_\mathrm{p}$) is calculated as an
interval between the same eruption time and the moment of the
minimum hourly Dst index for the given GMS.

To evaluate parameters of dimmings and arcades, we analyzed solar
images obtained in the 195~\AA\ channel of SOHO/EIT (dominating
line is Fe{\sc xii}, characteristic temperature is 1.3 MK). The
corresponding FITS files were downloaded from the EIT catalog
(\url{http://umbra.nascom.nasa.gov/eit/eit-catalog.html}).

In patrol CME watch observations, the 195~\AA\ images were
obtained usually with an imaging interval of 12 min. For the
present analysis, the solar rotation in the analyzed images was
compensated, and then the same fixed image before an eruption was
subtracted from each subsequent ones to obtain fixed-base images
\cite{ChertokGrechnev2005}. In most cases, a 3\,--\,4~h interval
from the eruption onset time was considered, \textit{i.e.}, a set
of 15\,--\,20 images was analyzed. During this time, the main
dimmings and arcades are already fully formed, but some minor
irrelevant evolutionary darkenings or brightenings appear on the
solar disk in this way.

Sometimes EIT observations with a 12-min imaging interval were
carried out in the 304~\AA\ channel instead of the 195~\AA\
channel. In this situation, when it was possible, we formed
difference images and evaluated parameters of dimmings and arcades
by using two or three suitable 195~\AA\ images obtained with a 6-h
interval. This was done, for example, for solar eruptions
corresponding to the CDAW events No.\,41 (eruption of 2001/08/14,
12:40), No.\,63 (double eruption of 2003/05/27, 23:07 and
2003/05/28, 00:27), and No.\,84 (eruption of 2005/05/26, 14:20).

Data processing was carried out with IDL employing SolarSoftware
general-purpose and instrument-specific routines as well as a
library and special software developed by the authors for the
present task. The whole package allows us to perform all necessary
procedures: calibrations of raw FITS files; compensation of the
solar rotation and subtraction of images; extraction of dimmings
and PE arcade which develop due to an analyzed CME; computation of
areas and total intensity within the dimming and arcade regions
according to chosen criteria; overlay of resulting images of the
dimmings and arcades with SOHO/MDI magnetograms and calculation of
the photospheric magnetic fluxes within these structures. In the
course of the analysis, thresholds of relative changes of
brightness were determined, which were optimal for evaluation of
parameters of the dimmings and arcades. Relative rather than
absolute thresholds were chosen for several reasons. Just the
relative thresholds allow us to take into account significant
dimmings in structures, whose brightness was small before an
eruption. It is also possible to reduce the influence of temporal
variations of the EIT detector characteristics as well as changes
in the calibration procedures. Finally, relative thresholds make
it possible in future to apply the quantitative results of the
present analysis to data from other EUV telescopes, in particular,
the Atmospheric Imaging Assembly (AIA; \opencite{Lemen2012}) on
board the \textit{Solar Dynamic Observatory} (SDO).

Parameters of dimming were computed from the so-called `portrait',
which shows in a single image all dimmings appearing all over the
event. The `dimming portrait' is formed as a maximum depth of the
depression (\textit{i.e.}, the minimum brightness) in each pixel
over the whole fixed-base difference set (see
\opencite{ChertokGrechnev2005}). The analysis showed that the
brightness depression of more than 40\% was an optimal criterion
for extraction of relevant significant dimmings. At this
threshold, shallow, short-lived, widespread, diffuse dimmings,
particularly associated with coronal waves, are not caught,
whereas main core dimmings adjoining to the eruption center and
other deep dimmings are displayed. At lower threshold values, many
remote evolutionary dimmings not related to the eruption under
consideration appear in difference images, while at larger
thresholds, some significant dimmings located near the eruption
center and obviously related to the eruption can be missed.

For PE arcades, a criterion turned out to be appropriate which
extracted an area around the eruption center where the brightness
in the 195~\AA\ channel exceeded 5\% of the maximum one. As has
been known, the area of a PE arcade increases with time.
Therefore, to avoid ambiguity, extraction of a PE arcade was
performed in an image temporally close to the maximum of the EUV
flux from the selected area. Usually this time is close to the
peak time of a corresponding GOES soft X-ray flare or somewhat
later. In particular, for events related to filament eruptions
outside ARs, the area of the arcade was calculated at the peak
time of the soft X-ray emission. In events associated with large
eruptions occurring in ARs and accompanied by very intense flares,
for example, such as X-class ones, a strong scattered light and a
long-duration bright, wide saturation streak crossing the eruption
center appear in EIT images. In such cases, the nearest frame
after disappearance of the distortion was taken for extraction of
the PE arcade and measurement of its parameters.

A total (unsigned) magnetic flux within dimming areas and PE
arcades is the most comprehensive and suitable parameter for the
analysis, because the intensity of GMSs and FDs as well as the
transit times (as will be shown below) are largely determined by
the magnetic characteristics of CMEs/ICMEs and their solar
sources. This parameter is evaluated within the contours of
dimmings and arcades determined according to the above
quantitative criteria, and thus, in fact, also takes into account
their area and intensity. In the present study for each event, the
line-of-sight magnetic field at the photospheric level is
calculated from SOHO/MDI level 1.8 magnetograms recalibrated in
December 2008
(\url{http://soi.stanford.edu/magnetic/index5.html}). The
magnetograms were routinely produced with an interval of 96 min.
We rebinned the magnetograms as well as EIT images to $512 \times
512$ pixels (with averaging) and resized the magnetograms to the
resulting pixel size of EIT. These procedures serve to minimize
measurement uncertainties. The 1-min magnetograms were mainly used
(in 43 events), while in seven events we were forced to use 5-min
ones. Calculations of the eruptive magnetic flux from closest
1-min and 5-min magnetograms have demonstrated that the
differences for the considered events did not exceed several
percent. The reason is that we are dealing with powerful
eruptions, which produce dimmings and arcades in sufficiently
strong magnetic fields for which the field, and therefore
contributions from noises in either 1-min or 5-min magnetograms
are not significant. Additional measurement issues are addressed
in Section~\ref{S-measurement_issues}.

\begin{figure} 
  \centerline{\includegraphics[width=\textwidth]
   {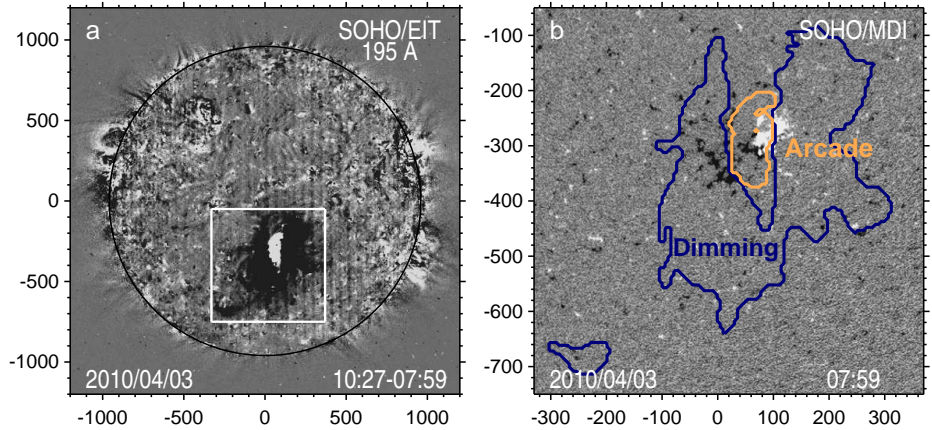}
  }
\caption{The 2010/04/03 eruption shown by the SOHO data: (a)~the
dimmings and arcade in the EIT 195~\AA\ fixed-base difference image;
(b)~an enlarged part of the MDI magnetogram corresponding to a
framed region in panel (a) with superposed dimming and arcade
contours, determined by the quantitative criteria, described in the
text.
    }
  \label{F-fig2}
  \end{figure}

To evaluate photospheric unsigned magnetic fluxes in dimmings
($\Phi_\mathrm{d}$) and arcades ($\Phi_\mathrm{a}$) as well as
their total flux ($\Phi = \Phi_\mathrm{d}+\Phi_\mathrm{a}$), we
take an MDI pre-event magnetogram closest to the eruption time and
compute the total magnetic fluxes within the corresponding regions
identified from EIT images. In this study, we use the total flux
of dimmings and arcade as a main parameter of an eruption.
Figure~\ref{F-fig2} illustrates the described procedures.

\subsection{Table}
 \label{S-table}

As a result of the procedures described above, the following
Table~\ref{T-table} of the analyzed events was formed. For each of
50 events, it starts with a GMS number corresponding to the CDAW
catalog (\citeauthor{Zhang2007a}
\citeyear{Zhang2007a,Zhang2007b}). Then information on the
geospace disturbance is provided including the GMS peak time,
minimum Dst value, FD magnitude, date and time of the disturbance
onset (SC). In column 6, the GMS type and identification level of
the solar eruptive source are given. The S1 and S2 codes mean the
separate storm caused by a single CME/ICME of a clear unambiguous
or only probable identification of a single eruption at the Sun,
respectively. The M2 code belongs to compound GMSs for which the
strongest decrease of the Dst index was selected and the most
powerful suitable solar eruption was determined as its probable
source of the identification level 2. The letter `R' after the S1,
S2, and M2 codes indicates that the solar source of the given GMS
was refined by us in comparison with the CDAW catalog as described
in Section~\ref{S-data_technique}. We will first examine the S1
group of single events with a reliable identification, and then
add the S2 and M2 events whose identification level is considered
as probable.

 \begin{kaprotate}
 \begin{table*}
 \caption{Parameters of major non-recurrent geomagnetic storms, Forbush decreases of the 23-rd cycle and their identified solar
 sources}
\label{T-table}
 \begin{tabular*}{\maxfloatwidth}{ccccccccccccc}
 \hline
 1 & 2 & 3 & 4 & 5 & 6 & 7 & 8 & 9 & 10 & 11 & 12 & 13 \\

 \multicolumn{1}{c}{} & \multicolumn{4}{c}{Geospace disturbance} & \multicolumn{1}{c}{} &
 \multicolumn{4}{c}{Eruption on the Sun} & \multicolumn{1}{c}{} & \multicolumn{2}{c}{Transit} \\

 \multicolumn{1}{c}{} & \multicolumn{4}{c}{} & \multicolumn{1}{c}{Identifi-} &
 \multicolumn{4}{c}{} & \multicolumn{1}{c}{} & \multicolumn{2}{c}{time [h]} \\

 & & & & Disturbance & cation & & Flare & & & Magnetic & & \\

CDAW & GMS peak & Dst & FD & onset, & level & Date, time & soft & Position & Type & flux $\Phi$ & $\Delta T_0$ & $\Delta T_\mathrm{p}$ \\

No. & (date, time) & [nT] & [\%] & Shock, SC & & & X-ray & & & [$10^{20}$ Mx] & & \\
 & & & & (date, time) & & & class & & & & & \\
 \hline
3 & 1997/05/15, 13 & $-$115 & 1.7 & 15, 01:59 & S1 & 12, 04:55 & C1.3 & N21W06 & AR & 95 & 69 & 80 \\
4 & 1997/10/11, 04 & $-$130 & 1.1 & 10, 16:12 & S1 & 06, $\sim$15:00 & --- & S27W05 & non-AR & 22 & 97 & 109 \\
5 & 1997/11/07, 05 & $-$110 & 2.1 & 06, 22:48 & S1 & 04, 05:58 & X2.1 & S14W33 & AR & 161 & 65 & 71 \\
9 & 1998/05/04, 06 & $-$205 & 3.5 & 04, 02:15 & M2 & 02, 13:42 & X1.1 & S15W15 & AR & 220 & 37 & 40 \\
16\,--\,17 & 1998/11/08, 07 & $-$149 & 7.4 & 08, 04:51 & S1, R & 05, 19:55 & M8.4 & N22W18 & AR & 276 & 57 & 59 \\
18 & 1998/11/13, 22 & $-$131 & 2.3 & 13, 01:43 & S1 & 09, 17:58 & C2.3 & N18E00 & non-AR & 76 & 80 & 100 \\
21 & 1999/09/23, 00 & $-$173 & 1.9 & 22, 12:09 & S1 & 20, 05:50 & C2.8 & S21W05 & non-AR & 47 & 54 & 66 \\
22 & 1999/10/22, 07 & $-$237 & 2.4 & 21, 02:25 & S1 & 17, 23:25 & C1.2 & S26E08 & non-AR & 64 & 75 & 104 \\
24 & 2000/02/12, 12 & $-$133 & 3.7 & 11, 23:52 & S1, R & 09, 20:06 & C7.4 & S17W40 & AR & 138 & 52 & 64 \\
26 & 2000/05/24, 09 & $-$147 & 3.3 & 23, 14:25 & M2 & 20, 05:35 & C7.6 & S15W08 & AR & 69 & 81 & 99 \\
27 & 2000/07/16, 01 & $-$301 & 11.7 & 15, 14:37 & S1 & 14, 10:24 & X5.7 & N22W07 & AR & 470 & 28 & 39 \\
29 & 2000/08/12, 10 & $-$235 & 2.7 & 11, 18:46 & S1 & 09, 16:22 & C2.3 & N11W11 & AR & 132 & 50 & 66 \\
30 & 2000/09/18, 00 & $-$201 & 8.1 & 17, 17:00 & M2 & 16, 04:26 & M5.9 & N14W07 & AR & 234 & 37 & 44 \\
32 & 2000/10/14, 15 & $-$107 & 3.6 & 12, 22:28 & S1 & 09, 23:43 & C6.7 & N01W14 & AR & 122 & 71 & 111 \\
34 & 2000/11/06, 22 & $-$159 & 7.8 & 06, 09:47 & S2, R  & 03, 19:02 & C3.2 & N02W02 & AR & 214 & 63 & 75 \\
35 & 2000/11/29, 14 & $-$119 & 2.7 & 28, 05:25 & M2 & 26, 16:48 & X4.0 & N18W38 & AR & 149 & 37 & 69 \\
36 & 2001/03/20, 14 & $-$149 & 2.9 & 19, 11:14 & S2, R & 15, 21:59 & C1.9 & N11W09 & AR & 108 & 85 & 112 \\
37 & 2001/03/31, 09 & $-$387 & 4.1 & 31, 00:52 & M2 & 29, 10:15 & X1.7 & N20W19 & AR & 377 & 39 & 47 \\
38 & 2001/04/12, 00 & $-$271 & 12.8 & 11, 13:43 & M2 & 10, 05:26 & X2.3 & S23W09 & AR & 294 & 32 & 43 \\
41 & 2001/08/17, 22 & $-$105 & 6.3 & 17, 11:03 & S1 & 14, 12:42 & C2.3 & N16W36 & non-AR & 68* & 70 & 81 \\
42 & 2001/09/26, 02 & $-$102 & 8.3 & 25, 20:25 & S1 & 24, 10:38 & X2.6 & S16E23 & AR & 271 & 34 & 39 \\
43 & 2001/10/01, 09 & $-$148 & 1.9 & 30, 19:24 & S1 & 28, 08:30 & M3.3 & N08E19 & AR & 170 & 59 & 72 \\
 \hline
 \end{tabular*}

 \end{table*}

\setcounter{table}{0}
\begin{table*}
 \caption{(\textit{Continued})}

 \begin{tabular*}{\maxfloatwidth}{ccccccccccccc}
 \hline
 1 & 2 & 3 & 4 & 5 & 6 & 7 & 8 & 9 & 10 & 11 & 12 & 13 \\

 \hline

44 & 2001/10/03, 15 & $-$166 & 2.5 & 03, 02:00 & S1 & 29, 11:06 & M1.8 & N13E03 & AR & 134 & 87 & 100 \\
45 & 2001/10/21, 22 & $-$187 & 5.4 & 21, 16:48 & S1 & 19, 16:30 & X1.6 & N15W29 & AR & 220 & 48 & 53 \\
46 & 2001/10/28, 12 & $-$157 & 1.9 & 28, 03:13 & M2 & 25, 15:02 & X1.3 & S18W19 & AR & 293 & 60 & 69 \\
48 & 2001/11/24, 17 & $-$221 & 9.2 & 24, 05:56 & M2 & 22, 23:30 & M9.9 & S14W36 & AR & 237 & 30 & 41 \\
50 & 2002/04/18, 08 & $-$127 & 6.2 & 17, 11:07 & S1 & 15, 03:55 & M1.2 & S15W01 & AR & 236 & 55 & 76 \\
51 & 2002/04/20, 09 & $-$149 & --- & 20, 00:00 & S1 & 17, 08:24 & M2.6 & S14W34 & AR & 286 & 64 & 73 \\
52 & 2002/05/11, 20 & $-$110 & 1.4 & 11, 10:14 & S1 & 08, 13:27 & C4.2 & S12W07 & AR & 137 & 69 & 79 \\
55 & 2002/08/21, 07 & $-$106 & 0.9 & 20, 14:00 & S2 & 16, 12:32 & M5.2 & S14E20 & AR & 117 & 97 & 114 \\
59 & 2002/10/04, 09 & $-$146 & 3.0 & 02, 23:00 & S1 & 30, 02:00 & C2.5 & S17W17 & non-AR & 20 & 93 & 127 \\
63 & 2003/05/30, 00 & $-$144 & 7.7 & 29, 12:24 & M2 & 27, 23:07 & X1.3 & S07W17 & AR & 160* & 37 & 49 \\
64 & 2003/06/18, 10 & $-$141 & 3.7 & 18, 05:01 & M2 & 14, $\sim$05:00 & --- & N22W15 & non-AR & 46 & 96 & 101 \\
66 & 2003/08/18, 16 & $-$148 & 2.6 & 17, 14:21 & S1, R & 14, 18:38 & C3.1 & S14E00 & AR & 177 & 68 & 93 \\
67 & 2003/10/30, 01 & $-$353 & 28.0 & 29, 06:11 & S1 & 28,11:10 & X17.2 & S16E08 & AR & 871 & 19 & 38 \\
68 & 2003/10/30, 23 & $-$383 & --- & 30, 16:00 & S1 & 29, 20:49 & X10.0 & S15W02 & AR & 520 & 19 & 26 \\
69 & 2003/11/20, 21 & $-$422 & 4.7 & 20, 08:03 & S1 & 18, 08:31 & M3.9 & N00E18 & AR & 133 & 48 & 60 \\
70 & 2004/01/22, 14 & $-$130 & 8.6 & 22,01:37 & S1 & 20, 00:40 & C5.5 & S13W11 & AR & 273 & 49 & 61 \\
73 & 2004/07/23, 03 & $-$99 & 4.3 & 22, 10:36 & S1 & 20, 12:32 & M8.6 & N10E35 & AR & 172 & 46 & 62 \\
74 & 2004/07/25, 17 & $-$136 & 4.6 & 24, 06:13 & S2, R & 22, 22:58 & M1.6 & N05E04 & AR & 257 & 31 & 66 \\
75 & 2004/07/27, 14 & $-$170 & 13.5 & 26, 22:49 & S1 & 25, 13:49 & M2.2 & N08W33 & AR & 363 & 33 & 48 \\
76 & 2004/08/30, 23 & $-$129 & 0.7 & 29, 10:06 & S2, R & 26, 13:04 & B8.4 & S11W38 & AR & 46.4 & 69 & 106 \\
77 & 2004/11/08, 07 & $-$374 & 5.2 & 07, 18:27 & M2, R & 06,00:34 & M3.9 & N09E05 & AR & 252 & 42 & 54 \\
78 & 2004/11/10, 11 & $-$263 & 8.3 & 09, 19:00 & M2, R & 07, 16:06 & X2.0 & N09W17 & AR & 290 & 51 & 67 \\
79 & 2005/01/18, 09 & $-$103 & 11.8 & 18, 06:00 & M2, R & 17, 09:52 & X3.8 & N15W25 & AR & 378 & 20 & 23 \\
81 & 2005/05/08, 19 & $-$110 & 5.1 & 08, 08:00 & M2, R & 06, 17:05 & C8.5 & S09E28 & AR & 294 & 39 & 50 \\
82 & 2005/05/15, 09 & $-$247 & 9.5 & 15, 02:38 & S1 & 13, 16:57 & M8.0 & N12E12 & AR & 266 & 34 & 40 \\
83 & 2005/05/20, 09 & $-$83 & 1.1 & 20, 04:01 & S1 & 16, 13:01 & C1.2 & N13W29 & AR & 54 & 87 & 92 \\
84 & 2005/05/30, 14 & $-$113 & 4.3 & 29, 09:52 & S2 & 26, 14:20 & B7.5 & S12E13 & AR & 89* & 68 & 96 \\
90 & 2006/12/15, 08 & $-$162 & 8.6 & 14, 14:14 & S1 & 13, 02:40 & X3.4 & S06W24 & AR & 222 & 35 & 53 \\
 \hline
 \end{tabular*}
 \end{table*}
 \end{kaprotate}

Columns 7\,--\,10 contain information on the corresponding solar
eruptive source: date and time of the eruption determined mainly
from the peak time of the soft X-ray flare emission, its GOES
class, and the position of the eruption site. In column 10, the
label `AR' means that the eruption occurred in and around an AR,
while the label `non-AR' indicates events resulting from filament
eruptions outside ARs. The resulting values of the parameters
analyzed in this paper are presented in columns 11\,--\,13. Here
and afterwards the main parameter of an eruption, the total
magnetic flux in dimmings and arcades at the photospheric level
($\Phi$), is expressed in units of $10^{20}$ Mx (maxwell). In
three events marked by the asterisks, the dimming and arcade areas
and the corresponding magnetic fluxes were measured by using two
to three suitable 195~\AA\ images obtained with the 6-h interval.
The two last columns present the onset ($\Delta T_0$) and peak
($\Delta T_\mathrm{p}$) transit times calculated as described
above.

\section{Results}
 \label{S-results}
\subsection{Forbush Decreases}
 \label{S-forbush_decreases}
\begin{figure} 
  \centerline{\includegraphics[width=\textwidth]
   {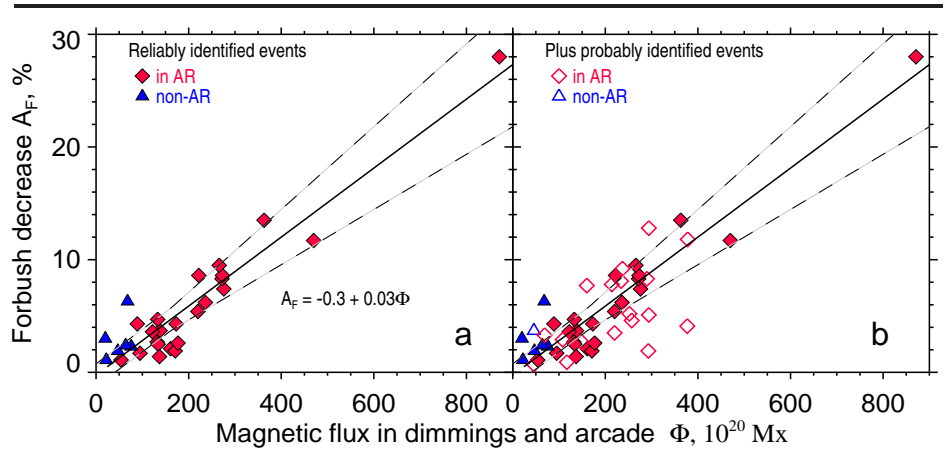}
  }
\caption{Dependence of the FD magnitude $A_\mathrm{F}$ on the
total magnetic flux in dimmings and arcades $\Phi$: a)~for single
geospace disturbances reliably identified with definite solar
eruptions (filled symbols); b)~for all considered events including
single and compound events with probable solar source
identification (open symbols). Here and afterwards the red
diamonds denote eruptions in ARs, and blue triangles denote
eruptions of quiescent filaments outside ARs. The dashed lines
delimit the accepted deviation band.
    }
  \label{F-fig3}
  \end{figure}

To assess how informative the total magnetic flux of dimmings and
arcades is and if it can really be used as a comprehensive
parameter of an eruption, we first of all examine how it is
related to the magnitude of FDs. Unlike GMSs, the magnitude of FDs
does not depend on the $B_z$ component being determined by the
magnetic field strength in a global ICME as well as its speed and
size. Figure~\ref{F-fig3}a shows the relationship between the
magnetic flux $\Phi$ and the FD magnitude $A_\mathrm{F}$ for
single geospace disturbances reliably identified with an
unambiguous solar eruption (the S1 group). Here and afterwards the
filled diamonds and triangles correspond to the AR and non-AR
eruptions, respectively. One can see that a conspicuous dependence
of the FD magnitude on the magnetic parameter of eruptions does
exist. On average, when the flux $\Phi$ increases from 30 to 900
(in $10^{20}$ Mx units), the FD amplitude $A_\mathrm{F}$ rises
from 0.8\% to 25\%. The dependence can be fitted with the
following linear regression relation
\begin{equation}
A_\mathrm{F}\ (\%) = -0.3 + 0.03 \Phi .
 \label{E-forbush}
\end{equation}
The correlation coefficient between $\Phi$ and $A_\mathrm{F}$
reaches $r \approx 0.94$. Note that this high correlation is only
marginally due to a great contribution from event No.\,67
(2003/10/30, 01) with the largest values of $\Phi$ and
$A_\mathrm{F}$ caused by the famous Halloween solar eruption on
2003/10/28. The high correlation persists even without this event.
For additional evaluation of scatter in data points, we accept a
deviation band bounded by $\pm 0.2$ of the regression line's slope
but not less than $\pm 1\%$ of $A_\mathrm{F}$. The latter
condition applies at relatively small eruptions, which correspond
to magnetic fluxes $\Phi \leq 180 \times 10^{20}$ Mx and small FD
magnitudes $A_\mathrm{F} \leq 5\%$. Calculations show that 18 out
of 29 events (\textit{i.e.} 62\%) fall into this deviation band.

The dependence of the FD magnitude on the eruption magnetic
parameter appears to be basically the same when single and
compound events with a probable solar source identification (the
S2 + M2 group, open symbols in Figure~\ref{F-fig3}b) are added to
single, unambiguously identified events. Here, as expected, the
scatter of points increases, and the correlation coefficient
somewhat reduces ($r \approx 0.86$). In this case, 22 points out
of 48 (\textit{i.e.}, 46\%) fall into the same deviation band.

From Figure~\ref{F-fig3} it is also visible that events associated
with filament eruptions outside ARs (triangles) are characterized
by relatively low values of magnetic fluxes $\Phi < 75 \times
10^{20}$ Mx. It is clear that this is caused by occurrence of such
eruptions in weak magnetic fields. Nevertheless, at least 3 out of
7 such events were accompanied by relatively strong FDs in the
range of 3\,--\,6.3\%. One possible reason for this unexpected
trend can be due to the fact that such non-AR filament eruptions
could lead to CMEs/ICMEs of sufficiently large size. As known, the
magnitude of FDs is determined not only by magnetic
characteristics of ICMEs, but also their global sizes. Additional
peculiarities of the non-AR events will be presented and discussed
below.

\subsection{Geomagnetic Storms}
 \label{S-geomagnetic_storms}

\begin{figure} 
  \centerline{\includegraphics[width=\textwidth]
   {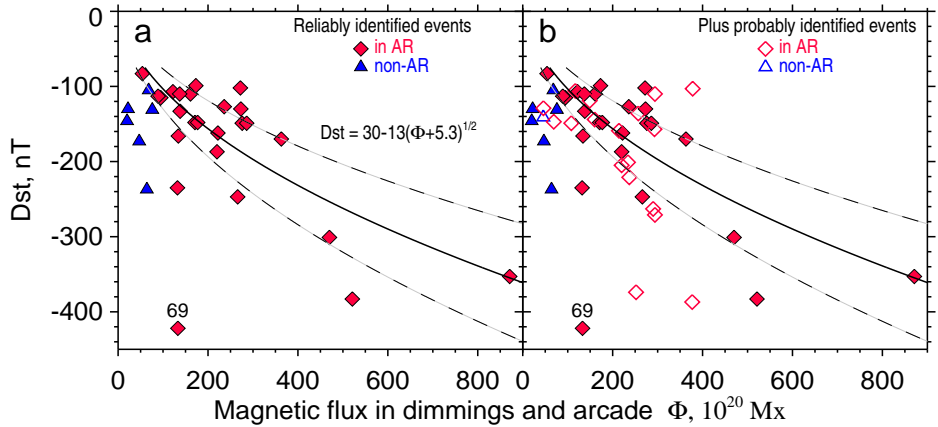}
  }
\caption{The same as in Figure~\ref{F-fig3} but for the dependence
of the geomagnetic storm intensity (Dst index) on the eruptive
magnetic flux $\Phi$.
    }
  \label{F-fig4}
  \end{figure}

The preceding section where FDs were considered has demonstrated
that the magnetic flux in dimmings and arcades has a high
informative potential for the space weather diagnostics as a
parameter of eruptions. The fact that a sufficiently high
correlation is revealed between this eruption parameter and the FD
magnitude allows us to expect that this parameter will be closely
related also to the GMS magnitude, especially if factors
determining the $B_z$ component in ICMEs would be taken into
account. In this paper, we will verify the relation between the
same magnetic flux $\Phi$ in dimmings and arcades, on the one
hand, and the geomagnetic Dst index on the other hand -- at the
first step, in a simplest way, without taking into account the
factors determining $B_z$. Note all the events analyzed here have
a negative $B_z$ component. The analyzed relationship might be
also scattered by other factors, which we do not take into account
-- for example, the ratio of sizes of an ICME near the Earth and a
pre-eruption magnetic structure on the Sun. The results are
presented in Figure~\ref{F-fig4}.

First of all let us consider single geomagnetic storms which are
not only reliably identified with a definite solar source (the S1
group), but associated exactly with AR eruptions (diamonds in
Figure~\ref{F-fig4}a). One can see that even in this simplified
analysis (\textit{i.e.}, disregarding $B_z$, but note all events
studied here have a southward $B_z$ component), the relationship
between $\Phi$ and Dst is noticeable: eruptions with larger
magnetic flux result in stronger GMSs. Event No.\,69 is a
conspicuous outlier. This is the famous event, in which a moderate
eruption on 2003/11/18 with moderately fast CMEs, relatively weak
flares, and modest dimming/arcade magnetic fluxes resulted in the
most intense GMS of the 23rd solar cycle with Dst $\approx -422$
nT. Unusual features of this outstanding event were discussed in a
number of papers (see, \textit{e.g.}, \opencite{Gopalswamy2005};
\opencite{Schmieder2011}; \opencite{Marubashi2012}; and references
therein). However, causes of this super-storm after the
comparatively insignificant solar eruption are still unclear. A
new detailed multi-spectral analysis of solar and interplanetary
manifestations in this event made by some of us with co-authors by
involving observations, which were not considered previously,
seems to have progressed in understanding the problem (we intend
to present the results in future papers). Here it is reasonable to
note that a combination of a quite moderate FD ($A_\mathrm{F}
\approx 4.7\%$) and strongest GMS (Dst $\approx -422$ nT)
registered in this event indicates that the ICME arrived at the
Earth orbit in a form of a relatively small cloud. Remote and
\textit{in situ} interplanetary measurements confirm this
conjecture. These circumstances suggest that in this case the
magnetic cloud expanded weakly during its propagation from the Sun
to the Earth and, as a result, preserved a strong magnetic field
inside ($B \approx 52$ nT). An additional decisive favorable
factor for the occurrence of the super-storm was that the $B_z$
component in the ICME was nearly antiparallel to the Earth's
magnetic dipole, so that almost the whole unusually strong
magnetic field of the ICME interacted with the Earth's
magnetosphere.

Excluding event No.\,69, Figure~\ref{F-fig4}a shows that as the
magnetic flux increases from 75\,--\,100 to 800\,--\,900 (in
$10^{20}$ Mx units), the GMS enhances from Dst $\approx -100$ nT
to Dst $\approx -(350-400)$ nT. This dependence can be expressed
by the formula
\begin{equation}
\mathrm{Dst}\ (\mathrm{nT}) = 30-13 (\Phi+5.3)^{1/2} .
 \label{E-dst}
\end{equation}
In this case, the correlation coefficient between the observed Dst
and the values calculated from the formula is $r \approx 0.67$.
However, the scatter of the points on this $\Phi-$Dst plot is
large, probably because the factors determining the sign of $B_z$
are not taken into account. The $\pm20\%$ deviation band relative
to the Dst$(\Phi)$ dependency bounded by the dotted lines in
Figure~\ref{F-fig4}a, contains 12 out of 29 (\textit{i.e.}, 41\%)
of the S1 events. If one takes into account the exceptional event
No.\,69, the correlation worsens to $r \approx 0.53$.

Figure~\ref{F-fig4}a also shows that the S1 events associated with
filament eruptions outside ARs (triangles) display again an
unexpected behavior, as in the case of FDs
(Section~\ref{S-forbush_decreases}). In spite of small magnetic
fluxes in dimmings and arcades, such filament eruptions produced
relatively intense GMSs. In Figure~\ref{F-fig4}a, at least 5 out
of 6 points lie below the Dst$(\Phi)$ curve and outside the
accepted deviation band. In this case, such a deviation in the
direction of stronger GMSs cannot be accounted for by possible
large sizes of ICMEs, because, unlike FDs, the GMS intensity is
determined by local rather than global characteristics of
interplanetary clouds at the site of their interaction with the
Earth's magnetosphere. Perhaps this property of non-AR events is
due to a selection effect. One should keep in mind that we study
here the strongest GMSs of Dst $ <-100$ nT and their eruptive
sources. Therefore, only several most significant non-AR events
were included in our consideration. Moreover, in all of the
diverging non-AR events, the magnetic field vector in the ICMEs
measured near the Earth was pointed practically south similar to
event No.\,69 mentioned above that favored enhanced GMSs. It
should be added that in two of the most deviating events No.\,21
and No.\,22 (see Table~\ref{T-table}), the total magnetic field in
the ICMEs (25 and 35~G) and their $B_z$ components (23 and 31~G)
were rather strong. Perhaps these events were not purely non-AR
eruptions, and the corresponding flux ropes were anchored in
related ARs.

Turning to Figure~\ref{F-fig4}b, one can see that the general
dependence between the dimming/arcade magnetic flux $\Phi$ and GMS
index Dst preserves its original appearance if single,
unambiguously identified events (the S1 group, filled symbols) are
supplemented with single and compound events of a probable solar
source identification (the S2 + M2 group, open symbols). Naturally
that in this case the correlation between the observed GMS
intensity and the calculated one from Equation (\ref{E-dst}) is
less, $r \approx 0.57$, and still decreases to $r \approx 0.49$ if
the exceptional event No.\,69 is taken into account.

\subsection{Transit Times}
 \label{S-transit_times}
Now we consider how the total magnetic flux $\Phi$ in dimmings and
arcades is related to two temporal parameters of GMSs, the onset
($\Delta T_0$) and peak ($\Delta T_\mathrm{p}$) transit times. Let
us remind that $\Delta T_0$ is defined as an interval between the
CME eruption time, which we take as the peak time of an associated
soft X-ray burst, and the arrival time of the corresponding
interplanetary disturbance (a shock wave) to the Earth indicated
particularly by SSC, and $\Delta T_\mathrm{p}$ is calculated as an
interval between the same eruption time and the moment of the
minimum Dst index of a given GMS. We realize that the propagation
time of CMEs/ICMEs from the Sun to the Earth depends on many
factors (lifting features in the corona, characteristics of the
background solar wind, interaction with other interplanetary
disturbances, \textit{etc.}), and the GMS peak time is determined
not only by the ICME speed, but also by the magnetic field
distribution in an ICME, \textit{i.e.}, in which part of it (shock
sheath ahead of an ICME, frontal or trailing component within its
body) the enhanced negative $B_z$ field is embedded. Nevertheless,
by comparing the eruptive flux with transit times, we want to
study to what extent the transit times (and therefore the 1~AU
ICME transit speed) are determined by parameters of a solar
eruption.

\begin{figure} 
  \centerline{\includegraphics[width=\textwidth]
   {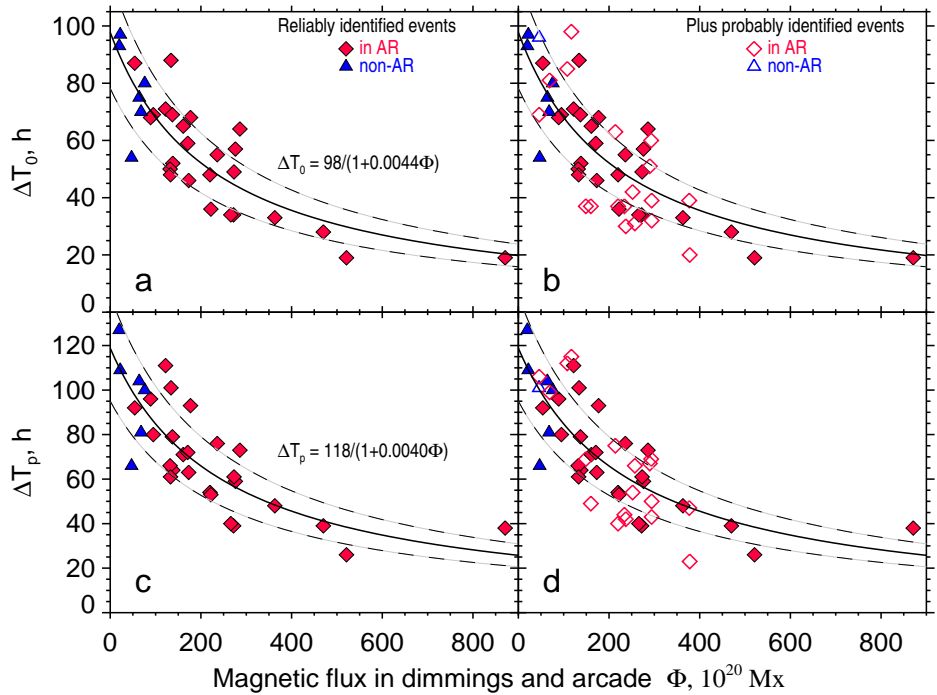}
  }
\caption{The same as in Figure~\ref{F-fig3} but for the dependence
of the onset ($\Delta T_0$, panels a,b) and peak ($\Delta
T_\mathrm{p}$, panels c,d) transit times on the eruptive magnetic
flux $\Phi$.
    }
  \label{F-fig5}
  \end{figure}

In Figure~\ref{F-fig5}a, the relationship between the eruptive
magnetic flux $\Phi$ and the onset transit time $\Delta T_0$ is
presented for single reliably identified S1 events including both
AR (diamonds) and non-AR (triangles) eruptions. The dependence
between $\Phi$ and $\Delta T_0$ is evident. The greater eruptive
magnetic flux (\textit{i.e.}, the more powerful eruption), the
shorter the transit time of the ICME-driven shock propagation from
the Sun to the Earth is, and the faster a GMS starts. For weak
magnetic fluxes $\Phi < 100 \times 10^{20}$ Mx, in most cases the
onset transit time is $\Delta T_0 \approx 70-95$~h, and for the
strongest eruptions with $\Phi \approx (500-900) \times 10^{20}$
Mx, the onset transit time comes to a level of about $\Delta T_0
\approx 20$ h, which corresponds to the average 1 AU ICME transit
speed of about 2100~km~s$^{-1}$. Analytically this dependence is
expressed as follows
\begin{equation}
\Delta T_0\ (\mathrm{h}) = 98/(1+0.0044\Phi).
 \label{E-t0}
\end{equation}
For the whole set of the S1 events under consideration, the
correlation coefficient between the observed onset times and the
$\Delta T_0$ calculated from expression (\ref{E-t0}) is
sufficiently high, $r \approx 0.84$. The $\pm 20\%$ deviation band
between the dotted lines in Figure~\ref{F-fig5}a contains 21 out
of 31 (\textit{i.e.}, 68\%) of the events. There are no
significant exceptions on this $\Delta T_0-\Phi$ chart.

The dependence between $\Phi$ and $\Delta T_0$ including the
features described above for the S1 events (filled symbols)
remains valid when less reliably identified and compound S2 + M2
events (open symbols) are added into consideration in
Figure~\ref{F-fig5}b. Here the correlation coefficient between the
observed onset times and those calculated from Equation
(\ref{E-t0}) only slightly reduces to $r \approx 0.81$, but the
scatter somewhat increases, and the number of points within the
same $\pm 20\%$ deviation band decreases to 28 out of 50
(\textit{i.e.}, 56\%).

The determining role of a solar eruptive source influences the
ICME speed so strongly that the peak transit time $\Delta
T_\mathrm{p}$ also exhibits a similar clear dependence on the
magnetic flux in dimmings and arcade $\Phi$ in spite of the
interfering factors mentioned above. As Figures~\ref{F-fig5}c and
\ref{F-fig5}d show, a similar expression
\begin{equation}
\Delta T_\mathrm{p}\ (\mathrm{h}) = 118/(1+0.0040\Phi)
 \label{E-tp}
\end{equation}
revealed from consideration of reliably identified S1 events can
be used for description of this dependence. It can be seen that at
small magnetic fluxes $\Phi <100 \times 10^{20}$ Mx the majority
of GMSs has the peak transit time $\Delta T_\mathrm{p} \approx
80-130$ h, and for most powerful eruptions with $\Phi \approx
(500-900) \times 10^{20}$ Mx the GMS peak becomes $\Delta
T_\mathrm{p} \approx 20-40$ h. This means again that solar
eruptions from areas of relatively small (large) dimming/arcade
magnetic flux are accompanied by low-speed (high-speed) ICMEs,
respectively. The correlation coefficient between the observed GMS
peak times and those calculated from Equation (\ref{E-tp}) is
approximately the same, $r \approx 0.81$, for both the S1 events
(Figure~\ref{F-fig5}c) and the whole set of the events
(Figure~\ref{F-fig5}d). In these diagrams, the relative number of
points within the $\pm 20\%$ deviation band is 65\% (20 out of 31)
and 58\% (29 out of 50) for the S1 and S1 + S2 + M2 event groups,
respectively. As for non-AR events (triangles in
Figure~\ref{F-fig5}), their majority shows the same pattern as
AR-associated ones, and, with their relatively smaller magnetic
fluxes, they show largest onset and peak transit times.

\section{Summary and Discussion}
 \label{S-discussion}
\subsection{Summary of Results}
We have studied relationships between characteristics of large
non-recurrent space weather disturbances of the 23rd solar cycle
in the form of GMSs with Dst $< -100$ nT and associated FDs, on
the one hand, and quantitative parameters of their solar source
manifestations such as EUV dimmings and PE arcades accompanying
the corresponding CMEs, on the other hand. In particular, the
total magnetic flux of the line-of-sight magnetic field at the
photospheric level within the dimming and arcade areas is used as
a main parameter of eruptions. The results presented above reveal
that, when a southward $B_z$ component is present, parameters of
space weather disturbances caused by CMEs/ICMEs are largely
determined by the power of solar eruptions (in terms of the total
magnetic flux in dimmings and arcades) in spite of many other
factors affecting the propagation of interplanetary disturbances
from the Sun to the Earth. This is true especially for eruptions
with a large magnetic flux. Just thanks to this fact, we were able
to establish the close statistical relationships of the magnetic
flux $\Phi$ in dimmings and arcades with the depth of FDs and
transit times as well as its correlation with GMSs initiated by
solar eruptions from the central part of the solar disk.
 \begin{itemize}
 \item
 First of all, to test the informative potential of the magnetic
flux as a parameter of an eruption, we analyzed its relationship
with the FD magnitude $A_\mathrm{F}$, because the latter, unlike
GMSs, does not depend on the $B_z$ component being determined by
the magnetic field strength in a global ICME as well as its speed
and size. It turned out that with an increase of the erupting
magnetic flux up to $900 \times 10^{20}$ Mx, the magnitude of the
corresponding FD grows linearly up to 25\%.
 \item
The above positive result allowed us to study the correlation of
the same eruptive parameter $\Phi$ with the GMS magnitudes, at the
first step, without taking into account the factors determining
$B_z$ near the Earth. We found that even in such a simplified
approach, the dependence between $\Phi$ and the geomagnetic Dst
index does exist indeed (all events considered in this study
contained a negative $B_z$ component). Stronger solar eruptions
characterized by larger magnetic fluxes result in more intense
GMSs up to Dst $\approx -400$~nT.
 \item
The same magnetic fluxes of dimmings and arcades exhibit a close
inverse correlation with the onset ($\Delta T_0$) and peak
($\Delta T_\mathrm{p}$) transit times measured, respectively, as
the propagation time of an ICME-driven shock from the Sun to the
Earth (a GMS onset) and the GMS peak time. With an increase of the
magnetic flux, both $\Delta T_0$ and $\Delta T_\mathrm{p}$ shorten
from 3\,--\,5 days to approximately 1~day. Equations (\ref{E-t0})
and (\ref{E-tp}) show that, for the first approximation, the
CME/ICME speed linearly increases with the strengthening of the
total magnetic flux in its solar source. On the other hand, we
have established that the FD and GMS magnitudes do depend on the
same eruptive magnetic flux, when a negative $B_z$ component is
present in those events. Juxtaposition of these relations sheds
light on the statistical correspondence of the magnetic field in
an ICME near the Earth on the near-the-Sun CME speed
\cite{Yurchyshyn2004} as well as a pronounced dependence of the FD
magnitude on the ICME speed \cite{Belov2009,RichardsonCane2011}.
 \item
 The physical meaning of Equations (\ref{E-t0}) and
(\ref{E-tp}) becomes clearer if we present them in a form $\Delta
T = R/(V_0 + k\Phi)$. Here $V_0$ is a velocity of the background
solar wind and $k\Phi$ is a CME/ICME velocity component governed
by parameters of a solar eruption. With $R = 1$~AU, we get $V_0
\approx 426$~km~s$^{-1}$, $k \approx 1.86$ for the onset transit
time $\Delta T_0$ (\ref{E-t0}) and $V_0 \approx 351$~km~s$^{-1}$,
$k \approx 1.41$ for the peak transit time $\Delta T_p$
(\ref{E-tp}), with $\Delta T$ being expressed in seconds. If the
eruptive component $k\Phi$ is small, then the arrival time of a
disturbance is mainly determined by the solar wind flow carrying
the ICME. In major events, $k\Phi \gg V_0$, the initial CME speed
determined by parameters of the eruption is high enough to ensure
the GMS onset time in 20\,--\,24~h despite the aerodynamic drag of
the solar wind. Note that \inlinecite{RichardsonCane2010} gave a
similar expression for the 1~AU transit speed, $V_\mathrm{tr}\
(\mathrm{km}\ \mathrm{s}^{-1}) = 400 + 0.8V_\mathrm{CME}$, where
the plane-of-the-sky CME speed stands instead of the magnetic
flux.
 \item
 The majority of events under consideration was caused by AR
eruptions. A few events produced by filament eruptions outside ARs
and characterized by small eruptive magnetic flux had long transit
times, caused GMSs, which were not so strong, and modest FDs.
However, some of such non-AR eruptions resulted in relatively
intense GMSs and FDs in comparison with AR eruptions of the same
value of the magnetic flux. For FDs this feature can be due to
larger size of corresponding CMEs/ICMEs, but it is not suitable
for GMSs. The most probable reason for this feature is that the
adopted criteria of extraction of the dimming and arcade areas are
not fully appropriate for these non-AR eruptions because the
latter have weaker dimmings and PE arcades in comparison with
eruptions occurring in ARs.
 \end{itemize}
Combining the established dependencies of the GMS severity and
transit times on the eruptive magnetic flux (Figures \ref{F-fig4}
and \ref{F-fig5}), we conclude that weak GMSs are characterized
mainly by long transit times and, conversely, short transit times
are typical of most intense GMSs. This relation follows from
\textit{in situ} measurements, which show high plasma speeds
within ICMEs during severe GMSs. From the present study it becomes
clear that this is caused by the fact that both the GMS intensity
(when a southward $B_z$ component is present) and ICME speed are
largely determined by the strength and extent of solar eruptions
expressed in the magnetic flux of EUV dimming and arcade areas.
These circumstances along with results of \inlinecite{Qiu2007},
\inlinecite{Vrsnak2005}, and conclusions of other authors suggest
that eruptions with larger magnetic fluxes initiate not only
bigger flares, but also faster CMEs/ICMEs. In this respect it is
worth noting that the tendency of inverse correlation between the
GMS intensity and ICME transit time appears to be supported by
data on the largest historical GMSs (see
\opencite{CliverSvalgaard2004}). In particular, in the famous
Carrington event of 1859, the severest GMS with estimated Dst
$\lsim -850$ nT \cite{SiscoeCrooker2006} commenced as early as
17~h after the large solar flare.

The new results on quantitative relationships between the magnetic
fluxes in dimming and arcade, FD and GMS magnitudes, and ICME
transit times obtained in our analysis are consistent with
conclusions of several previous studies. Formation of the helical
(poloidal) component of a magnetic flux rope by flare reconnection
was quantitatively confirmed by \inlinecite{Qiu2007} in their
comparisons of reconnected magnetic flux with the ICME magnetic
flux for several AR events. The detailed quantitative
correspondence between the reconnected flux and the rate of energy
release in the course of a flare was found (\textit{e.g.},
\opencite{Miklenic2009}). A well-defined correlation between the
plane-of-the-sky CME speed and the importance of the associated
flare was established indeed (\textit{e.g.},
\opencite{Vrsnak2005}). \inlinecite{Yurchyshyn2005} presented
correlations between the projected CME speed, Dst, and ICME
transit time. Most of the listed studies were related to
flare-related events in active regions; on the other hand,
\inlinecite{Chertok2009} showed that processes in non-AR filament
eruptions were basically similar to flare-related eruptions in AR.
The differences of non-AR eruptions from AR eruptions were found
to be mainly due to different character and strength of the
photospheric magnetic fields underneath. The magnetic fields in
non-AR events are weaker, with opposite polarities chaotically
alternating on small spatial scales, while the sizes of non-AR
eruptive filaments are much larger than those in AR eruptions.
These factors probably determine different parameters (and,
possibly, particularities of scenarios) of the two kinds of
eruptions. For all these reasons, a causal relationship between an
eruptive flare, CME development, and ICME expansion must exist,
and a quantitative correspondence between their parameters is
expected. All of these parameters appear to be determined by the
eruptive magnetic flux, which is directly related to the primary
driver of the flare-CME phenomenon, the non-potential magnetic
field in the corona. The larger the reconnected/eruptive magnetic
flux, the more powerful eruption, the stronger flare, the faster
CME, and eventually, the deeper FD and severer GMS (if a negative
$B_z$ is present) with a shorter delay after eruption are
expected.

The dependencies outlined above are expressed in the analytical
form with empirical expressions (Equations
(\ref{E-forbush})\,--\,(\ref{E-tp})). They form a tentative tool
allowing one to make an early diagnosis of geoefficiency of solar
eruptions and to carry out a short-term forecasting of main
parameters of non-recurrent space weather disturbances, including
estimations of a probable GMS intensity (if a negative $B_z$ is
present). The latter were obtained by assuming that the
corresponding ICMEs contain the necessary southern $B_z$
component, as well as all the events analyzed here. Already at the
time close to the maximum of corresponding soft X-ray flares by
using the solar EUV images and magnetograms, one can evaluate the
magnetic flux in dimmings and arcade and tentatively estimate with
this tool the expected maximum value of the GMS intensity as well
as the onset and peak times and the FD magnitude in advance from
one day for strongest eruptions to four days for relatively weak
ones. It should be remembered only that the dependencies presented
above were obtained for sufficiently large eruptions, which
produced strong geomagnetic storms of Dst $<-100$~nT.

\subsection{Measurement Issues}
 \label{S-measurement_issues}
The major uncertainties of our results are most likely due to
insufficiently known, quantified, or simply missed factors related
to ICMEs (size, configuration, orientation, background solar wind,
\textit{etc.}; see, \textit{e.g.}, \opencite{RichardsonCane2011})
and circumstances of their encounter with the Earth
\cite{Marubashi2012}. In this section we comment on uncertainties
of measurements from solar data and possible ways of future
improvements. Errors appear in measurements from magnetograms and
identification of arcade and dimming regions.
 \begin{itemize}
 \item
We mainly considered eruptions near the central meridian.
Nevertheless, regions of our analysis in some events extended
rather far from the solar disk center that could decrease the
measured total magnetic fluxes. We do not apply a radialization
correction of magnetograms, because a CME involves magnetic fields
of unknown orientations, and the radialization of the observed
line-of-sight magnetic component in this case might not provide
correct estimates. However, the correction factors, which could be
routinely applied, do not significantly differ from unity for the
majority of events. For example, possible corrections for the
pronouncedly non-central 2010/04/03 event (Figure~\ref{F-fig2})
are 15\% for the area and 9.6\% for the magnetic field strength. A
promising way to evaluate the total flux more accurately is a
magnetic field extrapolation, which allows one to reconstruct the
whole magnetic field vector. On the other hand, a non-central
position of an eruption implies a non-central encounter of the
corresponding ICME with the Earth that makes adequacy of such
corrections for our task questionable. Moreover, dimmings located
far from an eruption center are usually diffuse and shallow and
therefore automatically excluded by our selection criterion. For
all these reasons, we do not apply the projection corrections.
 \item
Some SOHO/MDI magnetograms suffer from the `saturation' in sunspot
umbrae due to limitations of the on-board data processing
\cite{LiuNortonScherrer2007}. The maximum field strength can be
underestimated by $> 20\%$. This artifact can affect measurements
in strongest events, when flare arcades cross sunspot umbrae, and
distort surrounding magnetic fields extrapolated from such
magnetograms. We cannot reliably compensate for such artifacts.
 \item
The contribution from noise in both 1-min and 5-min magnetograms
to our results is reduced ($\lsim 10-15\%$), because we use
rebinned magnetograms, in which four original MDI pixels are
averaged. Also, we consider severe GMSs (and large associated FDs)
caused by powerful solar eruptions including sufficiently deep
quasi-stationary dimmings. Such dimmings develop in regions of
increased brightness, \textit{i.e.}, preferentially above
photospheric regions with enhanced magnetic fields like plages
(see also \opencite{ChertokGrechnev2005}), so that the relative
contribution from noises is less important than in weaker-field
quiet Sun's regions.
 \item
Our identification of arcades and dimmings by using relative
thresholds ensures homogeneity of measurements. Identification in
events associated with very bright flares was complicated by
scattered light and overexposure effects like bright streaks
crossing the eruption site. Considerable contributions from these
distortions to our results are not expected, because each of such
events was carefully processed interactively.
 \end{itemize}

\subsection{Tentative Diagnostic Tool}
As an experiment, a tentative short-term forecasting of space
weather disturbances by using the presented results of solar
eruption diagnostics was carried out in the IZMIRAN Center of
Space Weather Forecasting during 2010 when the whole set of SOHO
data was available. Eruptions of the current 24th cycle from the
central zone of the solar disk were considered. Judging from
parameters of dimmings and arcades, the majority of them was
relatively small and according to estimations should have resulted
in rather faint space weather disturbances, and this was really
the case. One of the most significant eruptions of 2010 occurred
on 3 April in association with a B7.4 soft X-ray flare, which
peaked at 09:54 UT. The dimmings and arcade observed in this
eruption are shown in Figure~\ref{F-fig2}. Their total magnetic
flux in this case was $\Phi \approx 110 \times 10^{20}$ Mx. The
estimated FD magnitude $A_\mathrm{F} \approx 3\%$ and probable
maximum GMS intensity Dst $\approx -110$ nT corresponded to such
an eruptive magnetic flux. The actually observed $A_\mathrm{F}
\approx 2.9\%$ was close to the expected FD value, but the
observed Dst $\approx -73$ nT turned out to be somewhat weaker
than the estimated GMS intensity indicated above. Such a
combination of the FD and GMS values is possible when the
southward $B_z$ component of the ICME magnetic field comprises
only a part of the total magnetic field in an ICME. Data of the
OMNI catalog (\url{http://omniweb.gsfc.nasa.gov/}) reveal that
this was really the case: during the Dst peak, the southwards
component was about a half of the total field. As for the transit
times, the observed onset time $\Delta T_0 \approx 47$ h was
somewhat less than the estimated $\Delta T_0 \approx 66$ h, but
the observed peak time $\Delta T_\mathrm{p} \approx 77$ h was
close to the expected value $\Delta T_\mathrm{p} \approx 82$ h.
Approximately a similar correspondence between the estimated and
observed values was obtained in diagnostics of other sufficiently
large eruptions, which occurred in 2010 under the near-minimum
solar cycle conditions.

The described tentative tool based on calculations of the dimming
and arcade eruptive magnetic flux provides the earliest
diagnostics of the solar eruption geoeffectiveness and the
shortest lead time to forecast the maximum intensity, onset and
peak times of the forthcoming GMSs and FDs. We anticipate that
this tool would be used in future as a starting component of
combination of methods for short-term GMS and FD forecasting
including also those based on measurements of near-the-Sun CMEs,
MHD models, stereoscopic observations of ICME propagation, and
others (see Section~\ref{S-introduction}). A future real-time
forecasting thus could start just from near-solar-surface
manifestations of an earthward eruption and then specified as
additional data would come in the course of its expansion.

We consider the proposed tool as a preliminary one, because a
number of important issues should be addressed further for its
elaboration. First of all, the dependence of the Dst value on the
eruptive magnetic flux inferred in our study should be
complemented by taking account of the sign and strength of the
$B_z$ component in an ICME. This requires relating $B_z$ with
parameters of a solar source region. Further, for practical
application of the proposed quantitative diagnostic tool at the
present observational situation, it is necessary to develop
procedures to transit from EIT images and MDI magnetograms
obtained with SOHO during the 23th cycle to corresponding AIA
images and HMI magnetograms gathered at the present time with SDO
(see \opencite{Liu2012}).

\begin{acks}
We are grateful to an anonymous reviewer for useful remarks. The
authors thank the SOHO EIT, MDI, LASCO teams and the CDAW
participants for data and materials used in the present study.
SOHO is a project of international cooperation between ESA and
NASA. This research was supported by the Russian Foundation of
Basic Research under grants 09-02-00115, 11-02-00757, and
12-02-00037, the Program of basic research of the RAS Presidium
No.~22, and the Russian Ministry of Education and Science under
State Contract 16.518.11.7065.

\end{acks}

\end{article}

\end{document}